\documentclass[letterpaper, 10 pt, conference]{ieeeconf}
\IEEEoverridecommandlockouts

\usepackage{amsmath, amsthm, amssymb, amsfonts, mathtools, xargs, tensor, units, cite, stmaryrd, mathrsfs, algpseudocode, comment, graphicx}
\usepackage[unicode=true, bookmarks=true, bookmarksnumbered=false, bookmarksopen=false, breaklinks=false, pdfborder={0 0 1}, backref=false, colorlinks=false, hidelinks]{hyperref}
\makeatletter
\let\NAT@parse\undefined
\makeatother
\newcommand{\dif}{\mathop{}\!\mathrm{d}}
\usepackage[colorinlistoftodos]{todonotes}

\begin{document}
	\title{Online inverse reinforcement learning for nonlinear systems}
	
	\author{Ryan Self, Michael Harlan, and Rushikesh Kamalapurkar\thanks{The authors are with the School of Mechanical and Aerospace Engineering, Oklahoma State University, Stillwater, OK, USA. {\tt\small \{rself, michael.c.harlan, rushikesh.kamalapurkar\}@okstate.edu}.}}
	\maketitle
	\begin{abstract}
	This paper focuses on the development of an online inverse reinforcement learning (IRL) technique for a class of nonlinear systems. The developed approach utilizes observed state and input trajectories, and determines the unknown cost function and the unknown value function online. A parameter estimation technique is utilized to allow the developed IRL technique to determine the cost function weights in the presence of unknown dynamics. Simulation results are presented for a nonlinear system showing convergence of both unknown reward function weights and unknown dynamics. 
	\end{abstract}
	\section{Introduction}

	Human-robot interactions have become an important research
	area in the field of autonomous systems. Over the past decade, autonomous systems have been utilized to perform increasingly complex tasks, but their impact is ultimately limited by their inability to adapt to change. Humans can intuitively alter task objectives without being explicitly told, and can infer task objectives based on what other humans in their environment are trying to achieve. The ability of humans to detect the intent of others allows for better cooperation in teams. Similarly, the development of techniques to detect the intent of other entities will enable autonomous systems with the ability to detect subterfuge, to improve teamwork, and to learn appropriate responses to changing circumstances from the demonstrations.
	
	Based on the premise that the most succinct representation of the behavior of an entity is its reward structure, IRL aims to recover the reward (or cost) function by observing an agent performing a task and monitoring state and control trajectories of the observed agent. The IRL method developed in this paper learns the cost function and the value function of an agent under observation online, and in the presence of modeling uncertainties. 
	
	IRL methods were proposed in \cite{SCC.Ng.Russell2000} and cost function estimations using IRL were initially used for problems formulated as Markov Decision Processes (MDP) in \cite{SCC.Abbeel.Ng2004} -\cite{SCC.Ratliff.Bagnell.ea2006}. Since solutions to the IRL problem are generally not unique, the maximum entropy principle in \cite{SCC.Ziebart.Maas.ea2008} was developed to help differentiate between the various solutions. Beyond this, many extensions of IRL have been developed, including formulation of feature construction \cite{SCC.Levine.Popovic.ea2010}, solving IRL using gradient based methods \cite{SCC.Neu.Szepesvari2007},  and game theoretic approaches, as in \cite{SCC.Syed.Schapire2008}, which suggest the possibility of finding solutions that outperformed the expert. IRL was further extended to nonlinear problems using Gaussian Processes \cite{SCC.Levine.Popovic.ea2011}.
	
	There have been numerous techniques, such as IRL and inverse optimal control, that have been used to teach autonomous machines to perform specific  tasks in an offline setting, \cite{SCC.Mombaur.Truong.ea2010}. However, offline approaches to IRL cannot handle changes to task objectives or incorrect information in the problem formulation. The development of online IRL techniques would allow for the adaptation necessary for an autonomous system to reformulate the reward function based on the updated information. Recent work in \cite{SCC.Vamvoudakis.Lewis2010} - \cite{SCC.Wang.Liu.ea2016}, addresses reinforcement learning in real-time, and in \cite{SCC.Kamalapurkar.2018}, an online IRL method was developed for linear systems. The proposed method extends the results of \cite{SCC.Kamalapurkar.2018} to nonlinear systems.
	
	In the following, Section II details the mathematics notation used throughout the paper. Section III introduces the problem formulation. Section IV details an overview of the parameter estimation technique incorporated in the simulation. Section V explains the error metric used for the calculations. Section VI introduces the IRL algorithm. Section VII details how inaccurate data from the parameter estimation is removed. Section VIII is the analysis for convergence of the algorithm. Section IX shows the simulation for a nonlinear system and Section X is the conclusion of the paper.
	
	\section{Notation}\label{sec:Notation}
	The notation $\mathbb{R}^{n}$ represents the $n-$dimensional Euclidean space, and elements of $\mathbb{R}^{n}$ are interpreted as column vectors, where $\left(\cdot\right)^{T}$ denotes the vector transpose operator. The set of positive integers, not including 0, is denoted by $\mathbb{N}$. For $a\in\mathbb{R},$ $\mathbb{R}_{\geq a}$ denotes the interval $\left[a,\infty\right)$ and $\mathbb{R}_{>a}$ denotes the interval $\left(a,\infty\right)$. Unless otherwise specified, an interval is assumed to be right-open. If $a\in\mathbb{R}^{m}$ and $b\in\mathbb{R}^{n}$ then $\left[a;b\right]$ denotes the concatenated vector $\begin{bmatrix}a\\
	b
	\end{bmatrix}\in\mathbb{R}^{m+n}$. The notations and $\text{I}_{n}$ and $0_{n}$ denote  $n\times n$ identity matrix and the zero element of $\mathbb{R}^{n}$, respectively. Whenever clear from the context, the subscript $n$ is suppressed.

	\section{Problem formulation}\label{sec:NonLinearProblem}
	Consider an agent with the following nonlinear dynamics
	\begin{align}
	\dot{p}&=q\nonumber\\
	\dot{q}&=f(x,u)
	\end{align}
	where $p:\mathbb{R}^n\rightarrow \mathbb{R}$ is the position, $q:\mathbb{R}^n\rightarrow \mathbb{R}$ is the velocity, $x:=[p,q]^T$ is the state, and $u$ is the control. The dynamics in (1) can then be separated into 
	\begin{align}
	\dot{p}&=q\nonumber\\
	\dot{q}&= f^o(x,u) + g(x,u)
	\label{eq: NLSys}
	\end{align}
	where $f^o$ represents the known terms in the system dynamics, and $g$ represents the unknown terms. 
	
	The agent being observed, is trying to find the policy which minimizes the following performance index
	\begin{equation}
	J(x_0,u(\cdot)) = \int_{0}^{\infty}r(x(t;x_0,u(\cdot)),u(t))dt
	\end{equation}
	where $x(\cdot)$ is the trajectory of the agent generated by the optimal control trajectory $u(\cdot)$. To aid in cost function estimation, $r(x,u)$, can be parameterized as $r=W^T \sigma$, to be made precise in Section V, where $W$ represents the weights of the cost function to be approximated and $\sigma$ represent known continuously differential features. The objective of the proposed technique is for the system to identify the unknown weights of the cost function, $W$, along with identifying the unknown dynamics, $g(x,u)$.
	
	\section{Parameter estimator}\label{sec:SYSID}
	The parameter estimator developed by the authors in \cite{SCC.Kamalapurkar2017a} is utilized in this result. This section provides a brief overview of the same for completeness. For further details, the readers are directed to \cite{SCC.Kamalapurkar2017a}.
	
	To facilitate parameter estimation, let $ g(x,u) = \theta^T \sigma(x,u)+\epsilon(x,u) $, where  $\sigma:\mathbb{R}^n\times \mathbb{R}^m \rightarrow \mathbb{R}^n$ and $\epsilon:\mathbb{R}^n\times \mathbb{R}^m \rightarrow \mathbb{R}^p$ denote the basis vector and the approximation error, respectively, $\theta \in \mathbb{R}^{p \times n}$ is a constant matrix of unknown parameters. To obtain an error signal for parameter identification, the system in (1) is expressed in the form
	\begin{equation}
	\dot{q}\left(t\right)=f^{o}\left(x\left(t\right),u\left(t\right)\right)+\theta^{T}\sigma\left(x\left(t\right),u\left(t\right)\right)+\epsilon\left(x\left(t\right),u\left(t\right)\right).\label{eq:NLSPre Integral Form}
	\end{equation}
	Integrating (\ref{eq:NLSPre Integral Form}) over the interval $\left[t-\tau_{1},t\right]$
	for some constant $\tau_{1}\in\mathbb{R}_{>0}$ and then over the interval
	$\left[t-\tau_{2},t\right]$ for some constant $\tau_{2}\in\mathbb{R}_{>0}$,
	\begin{multline}
	\int_{t-\tau_{2}}^{t}\left(q\left(\lambda\right)-q\left(\lambda-\tau_{1}\right)\right)\dif\lambda=\int_{t-\tau_{2}}^{t}\int_{\lambda-\tau_{1}}^{\lambda}f^{o}\left(\gamma\right)\dif\gamma\dif\lambda\\+\theta^{T}\int_{t-\tau_{2}}^{t}\int_{\lambda-\tau_{1}}^{\lambda}\sigma\left(\gamma\right)\dif\gamma\dif\lambda+\int_{t-\tau_{2}}^{t}\int_{\lambda-\tau_{1}}^{\lambda}\epsilon\left(\gamma\right)\dif\gamma\dif\lambda,\label{eq:NLSDouble Integral Form}
	\end{multline}
	where the shorthands $f^o(\gamma)$, $\sigma(\gamma)$, and $\epsilon(\gamma)$ are used to denote $f^o(x(\gamma),u(\gamma))$, $\sigma(x(\gamma),u(\gamma))$, and $\epsilon(x(\gamma),u(\gamma))$, respectively.
	
	Using the Fundamental Theorem of Calculus and the fact that $q(t) = \dot{p}(t)$, the expression in (\ref{eq:NLSDouble Integral Form}) can be rearranged to form the affine system
		\begin{multline}
	P\left(t\right)=F\left(t\right)+\text{\ensuremath{\theta}}^{T}G\left(t\right)+E\left(t\right),\:\forall t\in\mathbb{R}_{\geq T_{0}}\label{eq:NLSDerivative Free Form}
	\end{multline}
	where 
	\begin{equation}
	P\left(t\right)\triangleq\begin{cases}
	\begin{gathered}p_1(t)\!-\!p_1\left(t\!-\!\tau_{1}\right)
	\!\\ \ \ \ -p_1\left(t\!-\!\tau_{2}\right)\!+\!\tau_{2},
	\end{gathered}
	& t\!\in\!\left[T_{0}\!+\!\tau_{1}\!+\!\tau_{2},\infty\right),\\
	0 & t<T_{0}+\tau_{1}+\tau_{2}
	\end{cases}\label{eq:NLSP}
	\end{equation}
	\begin{equation}
	F\left(t\right)\triangleq\begin{cases}
	\mathcal{I}f^{o}\left(t\right), & t\in\left[T_{0}+\tau_{1}+\tau_{2},\infty\right),\\
	0, & t<T_{0}+\tau_{1}+\tau_{2},
	\end{cases}\label{eq:NLSF}
	\end{equation}
	
	\begin{equation}
	G\left(t\right)\!\triangleq\!\begin{cases}
	\!\mathcal{I}\sigma\left(t\right), & t\!\in\!\left[T_{0}\!+\!\tau_{1}\!+\!\tau_{2},\infty\!\right),\\
	0 & t<T_{0}+\tau_{1}+\tau_{2},
	\end{cases}\label{eq:NLSG}
	\end{equation}
	and 
	\begin{equation}
	E\left(t\right)\triangleq\begin{cases}
	\mathcal{I}\epsilon\left(t\right), & t\in\left[T_{0}+\tau_{1}+\tau_{2},\infty\right),\\
	0 & t<T_{0}+\tau_{1}+\tau_{2}
	\end{cases}\label{eq:NLSU}
	\end{equation}
	
	For ease of exposition, it is assumed that a history stack, i.e., a set of ordered pairs $\left\{ \left(P_{i},{F}_{i},{G}_{i}\right)\right\} _{i=1}^{M}$ such that \begin{equation}
	P_{i}={F}_{i}+\text{\ensuremath{\theta}}^{T}{G}_{i}+\mathcal{E}_{i},\:\forall i\in\left\{ 1,\cdots,M\right\} ,\label{eq:NLSHistory Stack Compatibility}
	\end{equation} is available a priori, where $\mathcal{E}_i$ is a constant matrix. A history stack $\left\{ \left(P_{i},{F}_{i},{G}_{i}\right)\right\} _{i=1}^{M}$ is called \emph{full rank} if there exists a constant $\underline{c}\in\mathbb{R}$ such that \begin{equation} 0<\underline{c}<\lambda_{\min}\left\{ \mathscr{G}\right\} ,\label{eq:NLRank Condition} \end{equation} where the matrix  $\mathscr{G}\in\mathbb{R}^{p\times p}$ is defined as $\mathscr{G}\triangleq\sum_{i=1}^{M}{G}_{i}{G}_{i}^{T}$. The concurrent learning update law to estimate the unknown parameters is then given by
	\begin{equation}
	\dot{\hat{\theta}}\left(t\right)=k_{\theta}\Gamma\left(t\right)\sum_{i=1}^{M}{G}_{i}\left(P_{i}-{F}_{i}-\hat{\theta}^{T}\left(t\right){G}_{i}\right)^{T},\label{eq:NLSTheta Dynamics}
	\end{equation}
	where $k_{\theta}\in\mathbb{R}_{>0}$ is a constant adaptation gain and $\Gamma:\mathbb{R}_{\geq0}\to\mathbb{R}^{\left(2n^{2}+mn\right)\times\left(2n^{2}+mn\right)}$ is the least-squares gain updated using the update law
	\begin{equation}
	\dot{\Gamma}\left(t\right)=\beta_{1}\Gamma\left(t\right)-k_{\theta}\Gamma\left(t\right)\sum_{i=1}^{M}{G}_{i}{G}_{i}^{T}\Gamma\left(t\right).\label{eq:NLGamma Dynamics}
	\end{equation}
	Using arguments similar to \cite[Corollary 4.3.2]{SCC.Ioannou.Sun1996}, it can be shown that provided $\lambda_{\min}\left\{ \Gamma^{-1}\left(0\right)\right\} >0$, the least squares gain matrix satisfies
	\begin{equation}
	\underline{\Gamma}\text{I}_{p}\leq\Gamma\left(t\right)\leq\overline{\Gamma}\text{I}_{p},\label{eq:NLSStaFGammaBound}
	\end{equation}
	where $\underline{\Gamma}$ and $\overline{\Gamma}$ are positive
	constants, and $\text{I}_{n}$ denotes an $n\times n$ identity matrix.
	
	The affine error system in (\ref{eq:NLSDerivative Free Form}) motivates
	the adaptive estimation scheme that follows. The design is inspired
	by the \textit{concurrent learning} technique \cite{SCC.Chowdhary2010}.
	Concurrent learning enables parameter convergence in adaptive control
	by using stored data to update the parameter estimates. Traditionally,
	adaptive control methods guarantee parameter convergence only if the
	appropriate PE conditions are met \cite[Chapter 4]{SCC.Ioannou.Sun1996}.
	Concurrent learning uses stored data to soften the PE condition to
	an excitation condition over a finite time-interval.
	
	\section{Inverse Bellman Error}\label{sec:IBE}
	Under the premise that the observed agent makes optimal decisions, the Hamiltonian $H:\mathbb{R}^{2n}\times\mathbb{R}^{2n}\times \mathbb{R}^{m}\to\mathbb{R}$, which is defined as $H\left(x,p,u\right)\triangleq p^{T}f(x,u)+r\left(x,u\right)$, is convex in both the control signal $ u\left(\cdot\right) $, and the state, $ x\left(\cdot\right) $. Therefore, these satisfy the Hamilton-Jacobi-Bellman equation \begin{equation}
	H\left(x\left(t\right),\nabla_{x}\left(V^{*}\left(x\left(t\right)\right)\right)^{T},u\left(t\right)\right)=0,\forall t\in \mathbb{R}_{\geq 0},\label{eq:inverse HJB}
	\end{equation}
	where the unknown optimal value function is $ V^{*}:\mathbb{R}^{2n}\to \mathbb{R} $. The goal of IRL is to accurately estimate the cost function, $ r $, and to aid in the estimation of the cost function, let $ \hat{V}:\mathbb{R}^{2n}\times \mathbb{R}^{P} \to \mathbb{R}$, $ \left(x,\hat{W}_{V}\right)\mapsto \hat{W}_{V}^{T}\sigma_{V}\left(x\right) $ be the parameterized estimate of the optimal value function, $V^{*}$, where $ \hat{W}_{V}\in \mathbb{R}^{P} $ is the column of unknown reward weights, and $ \sigma_{V}:\mathbb{R}^{2n}\to\mathbb{R}^{P} $ are continuously differentiable features of the optimal value function which are known. Using $ \hat{\theta} $, $ \hat{W}_{V} $, $ \hat{W}_{Q} $, and $ \hat{W}_{R} $, which are the estimates of $ \theta $, $ W_{V}^{*} $, $ W_{Q}\coloneqq\left[q_{1},\cdots,q_{n}\right]^{T} $, and $ W_{R}\coloneqq\left[r_{1},\cdots,r_{m}\right]^{T} $, respectively, and the state, $ {x} $, in \eqref{eq:inverse HJB}, the inverse Bellman error $ \delta^{\prime}:\mathbb{R}^{2n}\times \mathbb{R}^{m}\times\mathbb{R}^{L+P+m}\times \mathbb{R}^{2n^{2}+mn}\to\mathbb{R} $ is obtained as \begin{align}
	\delta^{\prime}\left({x},u,\hat{W},\hat{\theta}\right)=&\hat{W}_{V}^{T}\nabla_{x}\sigma_{V}\left({x}\right) \ \hat{Y}({x},u)+\hat{W}_{Q}^{T}\sigma_{Q}\left({x}\right)\nonumber\\
	&+\hat{W}_{R}^{T}\sigma_{u}\left(u\right),
	\end{align}where $ \sigma_{u}\left(u\right)\coloneqq\left[u_{1}^{2},\cdots,u_{m}^{2}\right]$, $\hat{Y}({x},u)=\left[{x}_2;f^o({x},u)+\hat{g}({x},u)\right]$. Rearranging, we get \begin{equation}
	\delta^{\prime}\left({x},u,\hat{W}^{\prime},\hat{\theta}\right)=\left(\hat{W}^{\prime}\right)^{T}\sigma^{\prime}\left({x},u,\hat{\theta}\right),\label{eq:inverse BE}
	\end{equation}where $ \hat{W}^{\prime} \coloneqq \left[\hat{W}_{V};\hat{W}_{Q};\hat{W}_{R}\right] $, $ \sigma^{\prime}\left({x},u,\hat{\theta}\right)\coloneqq\left[\nabla_{x}\sigma_{V}\left({x}\right)\hat{Y}({x},u);\sigma_{Q}\left({x}\right);\sigma_{u}\left(u\right)\right] $.
	
	\section{Inverse Reinforcement Learning}\label{sec:IRL}
	Using the formulation of the inverse Bellman error in Section \ref{sec:IBE}, a history stack of the data can be gathered, and formulated into matrix form, resulting in 
	\begin{equation}
	\Delta^{\prime} = \hat{\Sigma}^{\prime} \hat{W}^{\prime},\label{homogeneous}
	\end{equation}where $ \Delta^{\prime}\coloneqq\left[\delta^{\prime}_{t}\left(t_{1}\right);\cdots;\delta^{\prime}_{t}\left(t_{N}\right)\right] $, $ \delta^{\prime}_{t}\left(t\right)\coloneqq \delta^{\prime}\left({x}\left(t\right),u\left(t\right),\hat{W}^{\prime},\hat{\theta}\left(t\right)\right)$, and $ \hat{\Sigma}^{\prime}\coloneqq\left[\left(\hat{\sigma}_{t}^{\prime}\right)^{T}\left(t_{1}\right);\cdots;\left(\hat{\sigma}_{t}^{\prime}\right)^{T}\left(t_{N}\right)\right] $. Using (\ref{homogeneous}), the IRL problem can be solved by finding the solution of the linear system for $\hat{W}$ that minimize the inverse Bellman error in \eqref{eq:inverse BE}. To facilitate the computation, the values of $ {x} $, $ u $, and $ \hat{\theta} $ are recorded at time instances $ \left\{t_{i}<t\right\}_{i=1}^{N} $ to generate the values $\left\{\hat{\sigma}_{t}^{\prime}\left(t_{i}\right)\right\}_{i=1}^{N}$, where $ N\in\mathbb{N} $, $ N>>L+P+m $, and  $ \hat{\sigma}_{t}^{\prime}\left(t\right)\coloneqq\sigma^{\prime}\left({x}\left(t\right),u\left(t\right),\hat{\theta}\left(t\right)\right) $. 
	
	Since the IRL problem is ill-posed, the solution is not unique. Taking a trivial solution, $\hat{W}'=0$, results in a minimal solution. This solution is not desired because it does not provide any useful information and should be discarded. However, in addition to this trivial solution, there are an infinite number of other solutions to the IRL problem. These solutions are constant multiples of the optimal solution. Meaning, the solutions to $r(x,u)$ and $a r(x,u)$, where $a$ is a positive constant, result in the exact same policy. 
	
	To deal with the issue of a non-unique solution, one of the reward function weights is assumed to be known. This mitigates the issue of the trivial solution, $\hat{W}'=0$ by forcing one of the weight to be nonzero, and removes the constant multiple solutions by fixing one known reward weight and, therefore, removes the scaling ambiguity. Taking $ \hat{W}_{R} $ to be known, the inverse BE in \eqref{eq:inverse BE} can then be expressed as \begin{equation}
	\delta^{\prime}\left({x},u,\hat{W},\hat{\theta}\right)=\hat{W}^{T}\sigma^{\prime\prime}\left({x},u,\hat{\theta}\right) + r_{1}\sigma_{u1}\left(u\right),
	\end{equation}where $ \sigma_{ui}\left(u\right) $ denotes the $ i $\textsubscript{th} element of the vector $ \sigma_{u}\left(u\right) $, the vector $ \sigma_{u}^{-} $ denotes $ \sigma_{u} $, with the first element removed, and $ \sigma^{\prime\prime}\left({x},u,\hat{\theta}\right)\coloneqq\left[\nabla_{x}\sigma_{V}\left({x}\right)\hat{Y}({x},u);\sigma_{Q}\left({x}\right);\sigma_{u}^{-}\left(u\right)\right] $.
	
	The closed-form nonlinear optimal controller corresponding to the reward structure in (3) provides the relationship\begin{equation}
	-2Ru\left(t\right)=\left(g^{\prime}\right)^{T}(\nabla_{x}\sigma_{V}\left(x\left(t\right)\right))^TW_{V}^{*}+\left(g^{\prime}\right)^{T}\nabla_{x}\epsilon\left(x\left(t\right)\right),
	\end{equation}which can be expressed as \begin{align*}
	-2r_{1}u_{1}\left(t\right) + \Delta_{u1}&=\sigma_{g1}\hat{W}_{V}\\
	\Delta_{u^{-}}&=\sigma_{g}^{-}\hat{W}_{V}+2\text{diag}\left(u_{2},\cdots,u_{m}\right)\hat{W}_{R}^{-},
	\end{align*}where $ \sigma_{g1} $ and $ \Delta_{u_{1}}$ denote the first rows and $ \sigma_{g}^{-}  $ and $ \Delta_{u^{-}} $ denote all but the first rows of $\sigma_{g}\coloneqq \left(g^{\prime}\right)^{T}(\nabla_{x}\sigma_{V}\left(x\right))^T $ and $ \Delta_u \coloneqq \left(g^{\prime}\right)^{T} \nabla_{x}\epsilon(x)$, respectively, and $ R^{-}\coloneqq\text{diag}\left(\left[r_{2},\cdots,r_{m}\right]\right) $. For simplification, let $ \sigma\coloneqq\left[\sigma^{\prime\prime},\, \begin{bmatrix}
	\sigma_{g}^{T}\\\Theta
	\end{bmatrix}\right]$, where \[ \Theta\coloneqq\left[0_{m\times 2n}, \,\,\begin{bmatrix}
	0_{1\times m-1}\\2\text{diag}\left(\left[u_{2},\cdots,u_{m}\right]\right)
	\end{bmatrix}\right]^{T} \]
	
	Updating the history stack in (\ref{homogeneous}), the formulation of removing a known reward weight element will generate the linear system \begin{equation}
	-\Sigma_{u1}=\hat{\Sigma}\hat{W}-\Delta^{\prime},
	\end{equation}where $ \hat{W} \coloneqq \left[\hat{W}_{V};\hat{W}_{Q};\hat{W}_{R}^{-}\right] $, $ \hat{\Sigma}\coloneqq\left[\hat{\sigma}_{t}^{T}\left(t_{1}\right);\cdots;\hat{\sigma}_{t}^{T}\left(t_{N}\right)\right] $, and $ \Sigma_{u1}\coloneqq\left[\sigma_{u1}^{\prime}\left(u\left(t_{1}\right)\right);\cdots;\sigma_{u1}^{\prime}\left(u\left(t_{N}\right)\right)\right] $, where $ \hat{\sigma}_{t}\left(\tau\right)\coloneqq\sigma\left({x}\left(\tau\right),u\left(\tau\right),\hat{\theta}\left(\tau\right)\right) $, $ \sigma_{u1}^{\prime}\coloneqq \left[r_1\sigma_{u1};2r_{1}u_{1};0_{\left(m-1\right)\times 1}\right]$, and the vector $ \hat{W}_{R}^{-} $ denotes $ \hat{W}_{R} $ with the first element removed.
	
	At any time instant $ t $, provided the history stack $ \mathcal{G}\left(t\right) $ satisfies \begin{equation}
	\textnormal{rank}\left(\hat{\Sigma}\right)=L+P+m-1,\label{eq:Rank Condition}
	\end{equation} then a least-squares estimate of the reward weights can be obtained as\begin{equation}
	\hat{W}\left(t\right)=-\left(\hat{\Sigma}^{T}\hat{\Sigma}\right)^{-1}\hat{\Sigma}^{T}\Sigma_{u1}.\label{eq:Least Squares}
	\end{equation} To improve numerical stability of the least-squares solution, the data recoded in the history stack is selected to maximize the condition number of $ \hat{\Sigma} $ while ensuring that the vector $ \Sigma_{u1} $ remains nonzero. The data selection algorithm is detailed in Fig. \ref{alg:DataSelect}.
\begin{figure}
	\hrulefill
	\begin{algorithmic}[1]
		\If{an observed, estimated or queried data point $ \left(x^{*},u^{*}\right) $ is available at $ t=t^{*} $}		
		\If{the history stack is not full}		
		\State add the data point to the history stack
		\ElsIf{$ \kappa\!\left(\!\!\left(\hat{\Sigma}\left(i\leftarrow*\right)\right)^{T}\!\!\left(\hat{\Sigma}\left(i\leftarrow*\right)\right)\!\!\right)\!<\!\xi_{1}\kappa\left(\hat{\Sigma}^{T}\hat{\Sigma}\right)$,\break\hspace*{1.2em} for some $ i $, and $\left\Vert \Sigma_{u1}\left(i\leftarrow*\right) \right\Vert\geq\xi_{2}$}		
		\State add the data point to the history stack
		\State $ \varpi \leftarrow 1 $
		\Else		
		\State discard the data point
		\State $ \varpi \leftarrow 0 $
		\EndIf
		\EndIf
	\end{algorithmic}
	\hrulefill
	
	\caption{\label{alg:DataSelect}Algorithm for selecting data for the history stack. The constants $\xi_{1}\geq0$ and $ \xi_{2}>0 $ are tunable thresholds. The operator $ \kappa\left(\cdot\right) $ denotes the condition number of a matrix. For the matrix $ \hat{\Sigma} = \left[\hat{\sigma}_{t}^{T}\left(t_{1}\right);\cdots;\hat{\sigma}_{t}^{T}\left(t_{i}\right);\cdots;\hat{\sigma}_{t}^{T}\left(t_{N}\right)\right] $, $ \Sigma\left(i\leftarrow*\right) \coloneqq \left[\hat{\sigma}_{t}^{T}\left(t_{1}\right);\cdots;\hat{\sigma}_{t}^{T}\left(t^{*}\right);\cdots;\hat{\sigma}_{t}^{T}\left(t_{N}\right)\right]$ and for the vector $ \Sigma_{u1} = \left[\sigma_{u1}\left(u\left(t_{1}\right)\right);\cdots;\sigma_{u1}\left(u\left(t_{i}\right)\right);\cdots;\sigma_{u1}\left(u\left(t_{N}\right)\right)\right] $, $ \Sigma_{u1}\left(i\leftarrow*\right) \coloneqq \left[\sigma_{u1}\left(u\left(t_{1}\right)\right);\cdots;\sigma_{u1}\left(u\left(t^{*}\right)\right);\cdots;\sigma_{u1}\left(u\left(t_{N}\right)\right)\right]$.}
\end{figure}

	\section{Purging to Exploit Improved Parameter Estimates}\label{sec:Purge}
	Due to the fact that $\hat{\Sigma}$ and $\Delta^{\prime}$ depends on the quality of the parameter estimates, a purging technique was incorporated in an attempt to remove inaccurate data. During the transient phase of the signals, the estimates in $ \mathcal{G} $ are likely to be less accurate, and therefore, the solution to the least squares problem will likely be poor. Therefore, a purging algorithm was developed and is detailed in Fig. \ref*{alg:Purging}.
	
	To determine the quality of the estimate, $\hat{\theta}$, a performance metric is sought. Using the dynamics in (\ref*{eq: NLSys}) and integrating over an interval $[t-T_1,t]$, a metric $q$ can be determined using the known state variable, $\dot{q}$, and the dynamics, $f^o(x,u)+\hat{\theta}^T\sigma(x,u)$. More specifically, 
	\begin{align}
	\bigg{|}q(t)-q(t-T_1)-\bigg{(}\int_{t-T_1}^{t}f^o(x(\tau),u(\tau))\nonumber\dif\tau\\+\int_{t-T_1}^{t}\hat{\theta}^T(\tau)\sigma(x(\tau),u(\tau))\dif\tau\bigg{)}\bigg{|}=\eta(t)
	\label{eq: ErrorMetric}
	\end{align} 
	If this is less than a predetermined constant threshold, then the estimate of $\hat{\theta}$ has improved and the data in the history stack should be purged. Since the error estimates exponentially converge to zero, a simpler time-based purging technique can also be incorporated, such as $|t-\eta|>\epsilon$, for a predetermined constant $\epsilon>0$, where $\eta $ is the time instant of the last purged event.
	
	An indicator variable, $ \eta $, as defined above, quantifies the quality of the current parameter estimates using a guess-and-check method, is used to purge and update the history stack. This updated history stack then updates the weight estimate $ \hat{W} $ according Fig. \ref{alg:Purging}. The algorithm is initialized with an empty history stack and an estimate for $ W_{0} $ is found. Using the algorithm in Fig. \ref{alg:DataSelect}, estimated values for $ {x} $, $ u $, $ \hat{\theta} $, and $ \eta $ are recorded in the history stack, where for $ t<T $, and $ \eta\left(t\right) $ is assumed to be infinite. The initial estimate of $ \hat{W} $ is kept constant until the history stack is full. Then, using $ \eqref{eq:Least Squares}, $ every time a new data point is added to the history stack, the weight estimate is updated.
	
	In addition to the data recorded along the trajectories of the demonstrator, a query-based approach can also be incorporated in IRL. In the query-based approach, a randomly selected state, $x_i$, is sent to the demonstrator and the corresponding control signal, $u_i$, is queried, and if these values are deemed to improve the estimate, then they are added to the history stack and used in the subsequent cost estimation calculations.
\begin{figure}
	\hrulefill
	\begin{algorithmic}[1]
		\State 	$ \hat{W}\left(0\right)\leftarrow W_{0} $, $ s\leftarrow 0 $
		\If{$\kappa\left(\hat{\Sigma}^{T}\hat{\Sigma}\right)<\underline{\kappa_{1}}$ and $ \varpi=1 $}		
		\State $ \hat{W}\left(t\right) \leftarrow -\left(\hat{\Sigma}^{T}\hat{\Sigma}\right)^{-1}\hat{\Sigma}^{T}\Sigma_{u1}$
		\Else
		\State Hold $ \hat{W} $ at the previous value
		\EndIf
		\If{$ \kappa\left(\hat{\Sigma}^{T}\hat{\Sigma}\right)<\underline{\kappa_{2}}$ and $ \eta\left(t\right)<\overline{\eta}\left(t\right) $}		
		\State empty the history stack
		\State $ s\leftarrow s+1 $
		\EndIf
	\end{algorithmic}
	\hrulefill
	\caption{\label{alg:Purging}Algorithm for updating the weights and the history stack. The constants $\underline{\kappa_{1}}>0$ and $ \underline{\kappa_{2}}>0 $ are tunable thresholds, the index $ s $ denotes the number of times the history stack was purged, and $ \overline{\eta}\left(t\right)\coloneqq\min\left\{\eta\left(t_{1}\right),\cdots,\eta\left(t_{M}\right)\right\} $.}
\end{figure}

	\section{Analysis}\label{sec:Ana}
	
	The parameter estimation used in the analysis is detailed in \cite{SCC.Kamalapurkar2017a}. To facilitate the analysis of the IRL algorithm, let $ \Sigma\coloneqq\left[\sigma\left(x\left(t_{1}\right),u\left(t_{1}\right),\theta\right);\cdots;\sigma\left(x\left(t_{M}\right),u\left(t_{M}\right),\theta\right)\right] $ and let $ \hat{W}^{*} $ denote the least-squares solution of $ \Sigma\hat{W}=-\Sigma_{u1} $. Furthermore, let $ W $ denote an appropriately scaled version of the ideal weights, i.e, $W\coloneqq \nicefrac{W}{r_{1}}$. Provided the rank condition in \eqref{eq:Rank Condition} is satisfied, the inverse HJB equation in (\ref{eq:inverse HJB}) implies that $ \Sigma W=-\Sigma_{u1}-E $, where $ E\coloneqq[\nabla_{x}\epsilon\left(x\left(t_{1}\right)\right)(f(x(t_1),u(t_1)))$; $\cdots$; $\nabla_{x}\epsilon\left(x\left(t_{M}\right)\right)(f(x(t_M),u(t_M)))] $. That is, $ \left\Vert  W+\left(\Sigma^{T}\Sigma\right)^{-1}\Sigma^{T}\Sigma_{u1} \right\Vert\leq\left\Vert \left(\Sigma^{T}\Sigma\right)^{-1}\Sigma^{T}E \right\Vert $. Since $ \hat{W}^{*} $ is a least squares solution, $ \left\Vert W-\hat{W}^{*} \right\Vert\leq\left\Vert \left(\Sigma^{T}\Sigma\right)^{-1}\Sigma^{T}E \right\Vert $.
	
	Let $ \hat{\Sigma}_{s} $, $ \Sigma_{u1_{s}} $, and $ \hat{W}_{s} $ denote the regression matrices and the weight estimates corresponding to the s\textsuperscript{th} history stack, respectively, and let $ \Sigma_{s} $ denote the ideal regression matrix where $ \hat{\theta}\left(t_{i}\right) $ in $ \hat{\Sigma}_{s} $ is replaced with the corresponding ideal value $ \theta $. Let $ \hat{W}^{*}_{s} $ denote the least-squares solution of $ \Sigma_{s}\hat{W}=-\Sigma_{u1_{s}} $. Provided  $ \hat{\Sigma}_{s} $ satisfies the rank condition in \eqref{eq:Rank Condition}, then $ \left\Vert W-\hat{W}^{*}_{s} \right\Vert\leq\left\Vert \left(\Sigma^{T}_{s}\Sigma_{s}\right)^{-1}\Sigma_{s}^{T}E \right\Vert $. Furthermore, $ \hat{W}_{s}-\hat{W}^{*}_{s} = \left(\left(\left(\hat{\Sigma}^{T}_{s}\hat{\Sigma}_{s}\right)^{-1}\hat{\Sigma}_{s}^{T}\right)-\left(\left(\Sigma^{T}_{s}\Sigma_{s}\right)^{-1}\Sigma_{s}^{T}\right)\right)\Sigma_{u1_{s}}$ Since the estimate $ \hat{\theta} $ exponentially converges to $ \theta $, the function $ \left(x,\theta\right)\mapsto \sigma\left(x,u,\theta\right) $ is continuous for all $ u $, and under the rank condition in \eqref{eq:Rank Condition}, the function $ \Sigma\mapsto\left(\Sigma^{T}\Sigma\right)^{-1}\Sigma^{T} $ is continuous, it can be concluded that $ \hat{W}_{s}\to\hat{W}^{*}_{s} $ as $ s\to\infty $, and hence, the error between the estimates $ \hat{W}_{s} $ and the ideal weights $ W $ is $ O\left(\overline{\epsilon}\right) $ as $ s\to\infty $.
	
	\section{Simulation}
	To verify the performance of the developed method, a nonlinear optimal control problem is selected with a known value function. The agent under observation has the following nonlinear dynamics\begin{align}
	\dot{x_1}&= x_2\nonumber\\
	\dot{x_2}&= f_1x_1 (\frac{\pi}{2}+\tan^{-1}(5x_1))+\frac{f_2x_1^2}{1+25x_1^2}+f_3x_2+3u
	\label{eq: NonLinear Simulation}
	\end{align}
	where the parameters, $f_1,f_2$ and $f_3$, are unknown constants to be estimated. The exact values of these parameters are $f_1=-1,f_2=-\frac{5}{2}$ and $f_3=4$.
	
	The cost function that the agent is trying to minimize is
	\begin{equation*}
	J(x_0,u(\cdot)) = \int_{0}^{\infty}(x_2^2+u^2)dt
	\end{equation*}
	resulting in the cost function weights to be estimated, $Q=diag(0,1)$ and $R=1$. The observed state and control trajectories, and a prerecorded history stack are used in the estimation of unknown parameters in the dynamics, along with the optimal value function parameters and the cost function weights.
	
	The closed form optimal nonlinear controller is
	\begin{equation*}
	u^*=-\frac{1}{2}R^{-1}g^T(x)(\nabla_xV)^T
	\end{equation*}
	resulting in the optimal controller
	\begin{equation*}
	u=-3x_2
	\end{equation*}
	and the optimal value function
	\begin{equation*}
	V^*=x_1^2(v_1+v_2\tan^{-1}(5x_1))+v_3x_2^2
	\end{equation*}
	resulting in the ideal function parameters $v_1=\frac{\pi}{2},$ $v_2=1$, and $v_3=1$.
	\begin{figure}
		\includegraphics[width=3.7in]{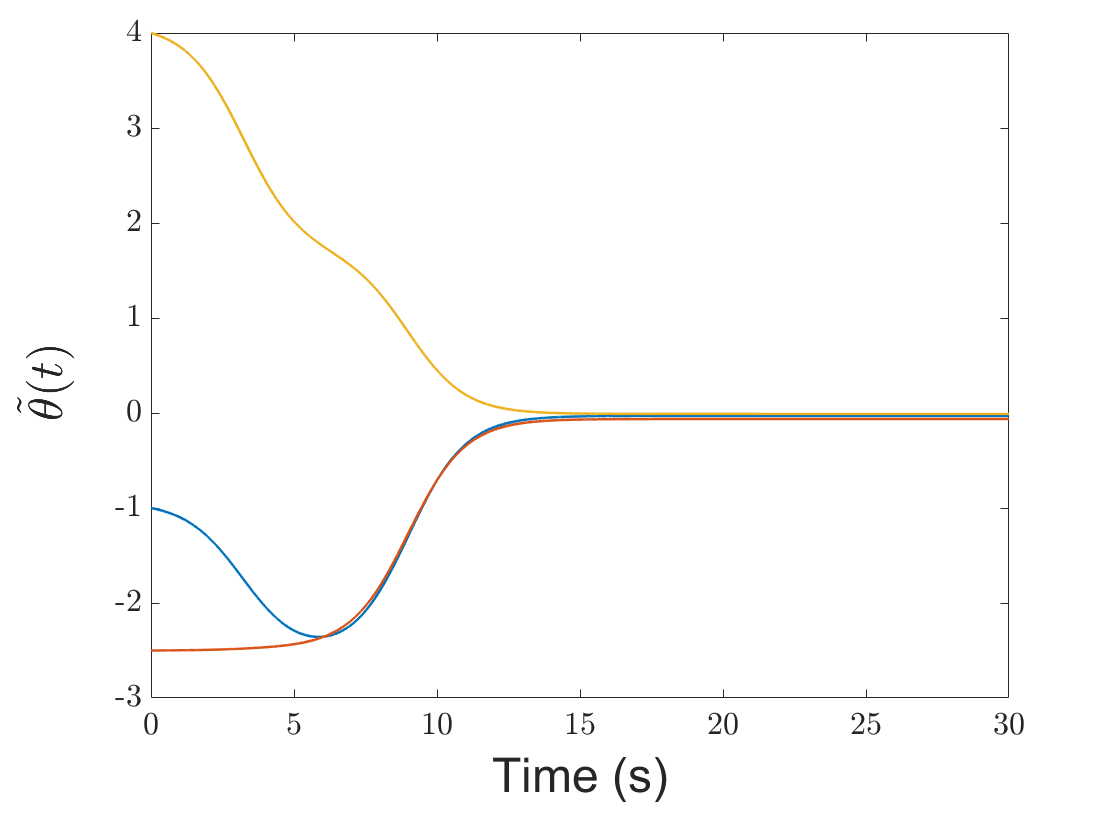}
		\caption{Estimation error for the unknown parameters in the system dynamics.}
		\label{fig: Parameter Estimation}
	\end{figure}
	\begin{figure}
		\includegraphics[width=3.7in]{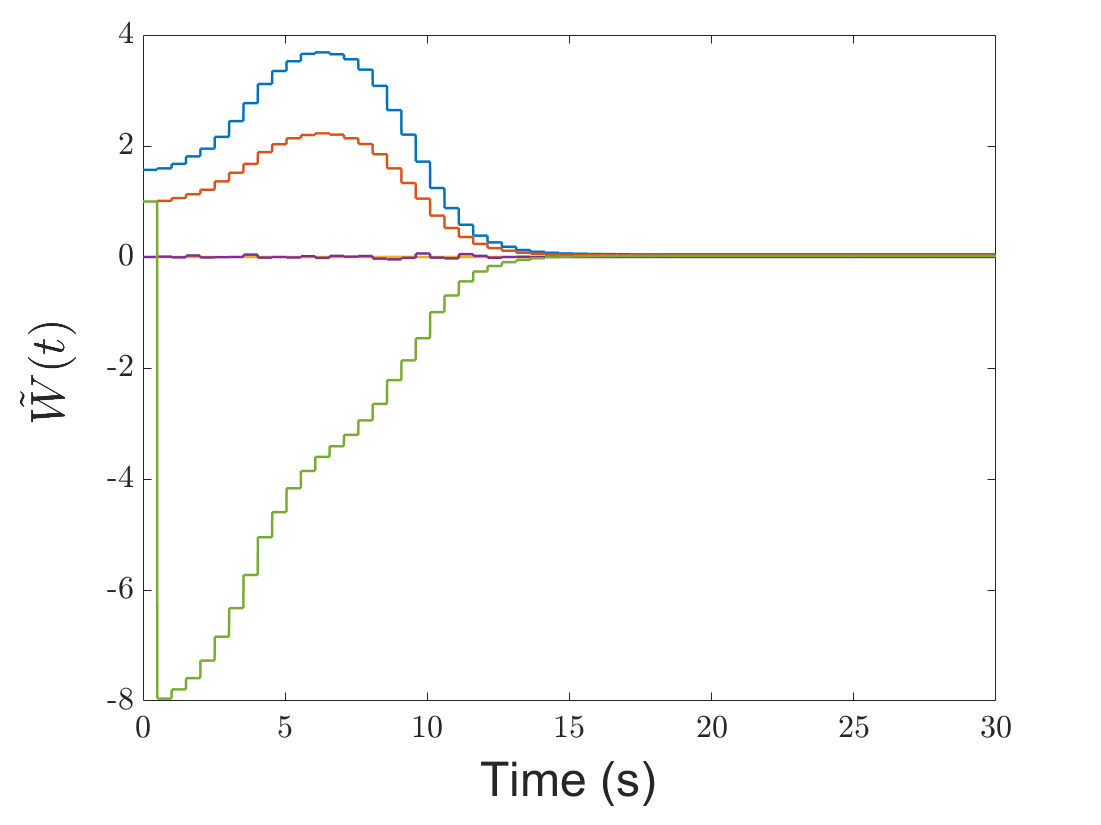}
		\caption{Estimation error for the unknown parameters in the cost function.}
		\label{fig: Cost Function Estimation}
	\end{figure}

	Figs. \ref{fig: Parameter Estimation} - \ref{fig: Cost Function Estimation} show the performance of the proposed method. Fig. \ref{fig: Parameter Estimation} shows the parameter convergence of the unknown part of the dynamics in (\ref{eq: NonLinear Simulation}). Fig. \ref{fig: Cost Function Estimation} shows the convergence of the unknown cost function weights, along with the unknown optimal value function parameters. The parameters used for the simulation were: $T_1=1$s, $T_2=0.6$s, $k_{\theta}=0.5/N$, $N=150$, $M=100$, $\beta_1=1$, $\Gamma(0)=\text{I}_{3\times3}$, and simulation time step, $T_s=0.005$s.
	\section{Conclusion}\label{sec:Con}
	In this paper, an online nonlinear inverse reinforcement learning method is developed. In order to facilitate cost function estimation online, a parameter estimator is incorporated to allow cost function estimation in the presence of unknown dynamics. Due to the dependency of the cost estimation method on the estimated dynamics, a purging technique is used to ensure that cost function estimation always utilizes the best available estimates of the system model. The method was validated by simulating a nonlinear system to show convergence of the unknown parameters in the dynamics, the cost function and the optimal value function.


\begin{thebibliography}{23}
	\bibitem{SCC.Ng.Russell2000}A. Y. Ng and S. Russell, "Algorithms for inverse reinforcement
	learning," in \textit{Proc. Int. Conf. Mach. Learn.} Morgan Kaufmann, 2000,
	pp. 663-670.
	
	\bibitem{SCC.Abbeel.Ng2004}Abbeel, Pieter, and Andrew Y. Ng. "Apprenticeship learning via inverse reinforcement learning." Proceedings of the twenty-first international conference on Machine learning. ACM, 2004.
	
	\bibitem{SCC.Ratliff.Bagnell.ea2006}N. D. Ratliff, J. A. Bagnell, and M. A. Zinkevich, "Maximum margin
	planning," in \textit{Proc. Int. Conf. Mach. Learn.}, 2006.
	
	\bibitem{SCC.Ziebart.Maas.ea2008}B. D. Ziebart, A. Maas, J. A. Bagnell, and A. K. Dey, "Maximum
	entropy inverse reinforcement learning," in \textit{Proc. AAAI Conf. Artif.
	Intel.}, 2008, pp. 1433-1438.
	
	\bibitem{SCC.Neu.Szepesvari2007}G. Neu and C. Szepesvari, "Apprenticeship learning using inverse
	reinforcement learning and gradient methods," in \textit{Proc. Anu. Conf.
	Uncertain. Artif. Intell.} Corvallis, Oregon: AUAI Press, 2007, pp.
	295-302.
	
	\bibitem{SCC.Syed.Schapire2008}U. Syed and R. E. Schapire, "A game-theoretic approach to
	apprenticeship learning," in \textit{Advances in Neural Information
	Processing Systems 20}, J. C. Platt, D. Koller, Y. Singer, and
	S. T. Roweis, Eds. Curran Associates, Inc., 2008, pp. 1449-1456.
	
	\bibitem{SCC.Levine.Popovic.ea2010}S. Levine, Z. Popovic, and V. Koltun, "Feature construction for
	inverse reinforcement learning," in \textit{Advances in Neural Information
	Processing Systems 23}, J. D. Lafferty, C. K. I. Williams, J. Shawe-
	Taylor, R. S. Zemel, and A. Culotta, Eds. Curran Associates, Inc.,
	2010, pp. 1342-1350.
	
	\bibitem{SCC.Levine.Popovic.ea2011}S. Levine, Z. Popovic, and V. Koltun, "Nonlinear inverse reinforcement
	learning with gaussian processes," in \textit{Advances in Neural Information
	Processing Systems 24}, J. Shawe-Taylor, R. S. Zemel, P. L. Bartlett,
	F. Pereira, and K. Q. Weinberger, Eds. Curran Associates, Inc.,
	2011, pp. 19-27.
	
	\bibitem{SCC.Mombaur.Truong.ea2010}K. Mombaur, A. Truong, and J.-P. Laumond, "From human to humanoid
	locomotion—an inverse optimal control approach," \textit{Auton.
	Robot.}, vol. 28, no. 3, pp. 369-383, 2010.
	
	\bibitem{SCC.Vamvoudakis.Lewis2010}K. G. Vamvoudakis and F. L. Lewis, "Online actor-critic algorithm to
	solve the continuous-time infinite horizon optimal control problem,"
	\textit{Automatica,} vol. 46, no. 5, pp. 878-888, 2010.
	
	\bibitem{SCC.Bian.Jiang.ea2014}T. Bian, Y. Jiang, and Z.-P. Jiang, "Adaptive dynamic programming
	and optimal control of nonlinear nonaffine systems," \textit{Automatica,}
	vol. 50, no. 10, pp. 2624-2632, 2014.
	
	\bibitem{SCC.Modares.Lewis2014}H. Modares and F. L. Lewis, "Optimal tracking control of nonlinear
	partially-unknown constrained-input systems using integral reinforcement
	learning," \textit{ Automatica,} vol. 50, no. 7, pp. 1780-1792, 2014.
	
	\bibitem{SCC.Kamalapurkar.Walters.ea2016}R. Kamalapurkar, P. Walters, and W. E. Dixon, "Modelbased
	reinforcement learning for approximate optimal regulation,"
	\textit{Automatica,} vol. 64, pp. 94-104, Feb. 2016.
	
	\bibitem{SCC.Wang.Liu.ea2016}D. Wang, D. Liu, H. Li, B. Luo, and H. Ma, "An approximate optimal
	control approach for robust stabilization of a class of discrete-time
	nonlinear systems with uncertainties," \textit{IEEE Trans. Syst. Man Cybern.
	Syst.}, vol. 46, no. 5, pp. 713-717, 2016.
	
	\bibitem{SCC.Ioannou.Sun1996}P. Ioannou and J. Sun, \textit{Robust adaptive control.} Prentice Hall, 1996.
	
	\bibitem{SCC.Xian.Queiroz.ea2004}B. Xian, M. S. de Queiroz, D. M. Dawson, and M. McIntyre, "A
	discontinuous output feedback controller and velocity observer for
	nonlinear mechanical systems," \textit{Automatica,} vol. 40, no. 4, pp. 695-
	700, 2004.
	
	\bibitem{SCC.Kamalapurkar2017a}Kamalapurkar, Rushikesh. "Simultaneous state and parameter estimation for second-order nonlinear systems." \textit{Decision and Control (CDC),} 2017 IEEE 56th Annual Conference on. IEEE, 2017.
	
	\bibitem{SCC.Chowdhary.Johnson2011a} G. Chowdhary and E. Johnson, "A singular value maximizing data
	recording algorithm for concurrent learning," in \textit{Proc. Am. Control Conf.,}
	2011, pp. 3547-3552.
	
	\bibitem{SCC.Kersting.Buss2014} S. Kersting and M. Buss, "Concurrent learning adaptive identification
	of piecewise affine systems," in \textit{Proc. IEEE Conf. Decis.} Control, Dec.
	2014, pp. 3930-3935.
	
	\bibitem{SCC.Parikh.Kamalapurkar.easubmitteda} A. Parikh, R. Kamalapurkar, and W. E. Dixon, "Integral concurrent
	learning: Adaptive control with parameter convergence without PE
	or state derivatives," submitted, see arXiv:1512.03464, iEEE Control
	Systems Letters.
	
	\bibitem{SCC.Kamalapurkar.Reish.ea2017} R. Kamalapurkar, B. Reish, G. Chowdhary, and W. E. Dixon,
	"Concurrent learning for parameter estimation using dynamic statederivative
	estimators," \textit{IEEE Trans. Autom. Control,} vol. 62, no. 7, pp.
	3594-3601, Jul. 2017.
	
	\bibitem{SCC.Chowdhary2010} G. Chowdhary, "Concurrent learning for convergence in adaptive control
	without persistency of excitation," Ph.D. dissertation, Georgia Institute
	of Technology, Dec. 2010.
	
	\bibitem{SCC.Kamalapurkar.2018} R. Kamalapurkar, "Inverse reinforcement learning in continous time and space," arXiv preprint arXiv:1801.07663, Jan. 2018.
\end{thebibliography}
\end{document}